\documentclass[twocolumn,showpacs,preprintnumbers,amsmath,amssymb]{revtex4}
\usepackage{psfig}
\usepackage{graphicx}% Include figure files
\usepackage{dcolumn}% Align table columns on decimal point
\usepackage{bm}% bold math

\begin{document}

\title{Exciton lifetime in InAs/GaAs quantum dot molecules}

\author{C. Bardot, M. Schwab, and M. Bayer}
\affiliation{Experimentelle Physik II, Universit\"at Dortmund,
D-44221 Dortmund, Germany}

\author{S. Fafard, Z. Wasilewski, and P. Hawrylak}
\affiliation {Institute for Microstructural Sciences, National
Research Council of Canada, Ottawa, K1A OR6, Canada}

\date{\today}

\begin{abstract}
The exciton lifetimes $T_1$ in arrays of InAs/GaAs vertically
coupled quantum dot pairs have been measured by time-resolved
photoluminescence. A considerable reduction of $T_1$ by up to a
factor of $\sim$ 2 has been observed as compared to a quantum dots
reference, reflecting the inter-dot coherence. Increase of the
molecular coupling strength leads to a systematic decrease of
$T_1$ with decreasing barrier width, as for wide barriers a
fraction of structures shows reduced coupling while for narrow
barriers all molecules appear to be well coupled. The coherent
excitons in the molecules gain the oscillator strength of the
excitons in the two separate quantum dots halving the exciton
lifetime. This superradiance effect contributes to the previously
observed increase of the homogeneous exciton linewidth, but is
weaker than the reduction of $T_2$. This shows that as compared to
the quantum dots reference pure dephasing becomes increasingly
important for the molecules.
\end{abstract}

 \maketitle

The development of high quality semiconductor quantum structures
of varying dimensionality has allowed attention to be focused on
coupling of these systems. Some interest in such activities arises
from application side as the resulting functional units have the
potential to form building blocks of a new generation of
electronic and optoelectronic devices. Some interest arises also
from basic physics as it gives detailed insight into quantum
mechanical coupling, for example. The most simple structure is a
pair of quantum dots (QDs), coupled by tunnelling. For their
fabrication, a variety of techniques may be applied such as double
cleaved edge overgrowth \cite{SchedelbeckSC97}, lateral patterning
of double quantum wells \cite{FujisawaSC98} or gating of
two-dimensional electron gases \cite{BlickPRL98}.

Here we focus on vertically coupled dot structures fabricated by
self-assembly, which seems ideally suited for their realization:
when growing two QD layers in close vicinity, the strain that
surrounds a dot in the first, lower lying layer enforces the
growth of a second dot on top of the first \cite{stacking}. While
vertical stacking is well established, spectroscopic studies of
these structures are just at their beginning
\cite{ShtrichmanPRB02}. Spectra taken on arrays suffer from
inhomogeneous broadening. Only recently it has become possible to
observe a well resolved confined shell structure in
photoluminescence of InAs/GaAs quantum dot molecules (QDMs). This
progress has been made possible by extending the conventional
Stransky-Krastanow growth scheme by an In-flush \cite{Fafard99}.
Through this modification the dot geometries in the two layers
become very similar, while the size of the upper dot is
considerably larger than that of the lower one in the non-extended
fabrication scheme. In particular, the homogeneity of the dot
height, which is the most crucial point for reducing the ensemble
broadening, is improved.

A proof of quantum mechanical coupling in these vertically aligned
QDs was provided by exciton fine structure studies: pronounced
anticrossings were observed in the magnetic field dispersion of
the fine structure because of field induced state hybridizations
\cite{OrtnerPRL03}. In addition, systematic dependencies of a
variety of spectroscopic quantities such as orbital and spin
energy splittings, diamagnetic shifts etc. on barrier width were
observed \cite{BayerSC01,OrtnerPRB05}. The entirety of these
results strongly support tunnelling as coupling mechanism in the
QDMs, through which spatial coherency of the wave function is
established.

Besides coherency in space, also temporal coherency of the
excitons has already been addressed in arrays of InAs/GaAs QDMs,
by measuring the dephasing with four wave mixing
\cite{BorriPRL03}. For the parameters by which dephasing in
three-dimensionally confined geometries can be characterized, such
as homogenous linewidth, zero-phonon line weight and activation
energies, systematic dependencies on barrier width have been
observed. Most importantly, the homogeneous linewidth at cryogenic
temperatures increases strongly with decreasing barrier width by a
factor of $\sim 6$ as compared to the QDs reference. In the dots
the dephasing time is about 600 ps, and it is reduced to slightly
more than 100 ps only for molecules with narrow 4 and 5 nm
barriers. For QDs it has been shown that the exciton dephasing is
ultimately limited by the radiative lifetime of the excitons, and
pure dephasing may not be important, depending on the dot
structure under study. \cite{LangbeinPRB04} One point of the
present manuscript is to assess the importance of pure dephasing
in molecule structures. For this purpose we complement the
previous four-wave-mixing studies by exciton lifetime
measurements. By doing so, we also address the effect of spatial
coherency of the exciton wavefunction on its oscillator strength.

Details of the QDM sample fabrication with barrier widths of $d =$
4, 5, 6, 7, 8, and 16 nm barriers can be found elsewhere.
\cite{Fafard99,nonanneal} The QDM samples were mounted on the cold
finger of a microscope cryostat, which allowed a variation of the
sample temperature down to 5 K. For optical excitation a
mode-locked Ti:sapphire laser pumped by a frequency doubled
Nd:YVO$_4$ laser was used. The system provides trains of pulses
with a duration of $\approx$ 150 fs at a repetition rate of 76
MHz. The laser wavelength was 780 nm, corresponding to an
excitation energy slightly above the GaAs barrier. Also
experiments with excitation wavelengths around 850 nm
(energetically below the GaAs barrier into the wetting layer) were
performed, but no significant changes of exciton dynamics were
observed. The laser was focussed to a spot size of 10 $\mu$m. From
the structural density of $\sim 10^{10}$ cm$^{-2}$ we estimate
that a few thousand molecule structures were addressed
spectroscopically. The excitation power was 40 Wcm$^{-2}$, small
enough to avoid many body effects and to address single excitons.
The emission was spectrally analyzed by a single grating
monochromator ($f$~=~0.5 m, 300 rules per mm) and detected by a
Hamamatsu synchroscan streak-camera equipped with a Peltier-cooled
S1-tube. To enhance sensitivity it contains a double multichannel
plate. With this configuration a time resolution of about 30 ps
was obtained. \cite{singleQDM}

\begin{figure}
  \centering
  \centerline{\psfig{figure=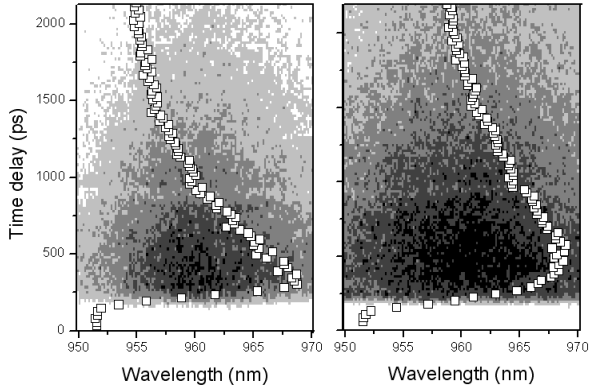,width=\columnwidth}}
\caption{Grey scale contour plots of the photoluminescence
emission decay from the QDs reference sample (the right panel) and
the QDM sample with a 4 nm wide barrier (the left panel) at $T$ =
10 K. The signal was integrated in both cases for 60 s. Symbols
are characteristic traces at the center of the emission band. The
peak intensity has been normalized to unity. Note that there is
some intensity reduction around 900 ps, due to reduced sensitivity
of the streak screen. The data are raw data have not been
corrected for this, also in the following.} \label{streakscreens}
\end{figure}

Fig. \ref{streakscreens} shows streak camera screens recorded at
$T$ = 10 K for the QDs reference (the right panel) and the $d$ = 4
nm barrier QDMs (the left panel): normalized contour plots of the
photoluminescence emission as function of the time delay after
pulsed excitation (the vertical axis) and the emission wavelength
(the bottom axis) are given. From this comparison, the
luminescence decay time in the molecules is much shorter than in
the QDs. We also find that within the inhomogeneously broadened
emission band in both cases the decay time does not vary strongly.
For the 4 nm barrier QDM sample the average tunnel splitting is
larger than 30 meV and the excitons relax fast into the ground
state, so that only emission from this level is detected. The
lifetimes are about the same within the luminescence band, showing
that the variation of exciton oscillator strength with emission
energy is small.

Excitonic luminescence from both tunnel split exciton states is
observed for tunnel splittings smaller than 30 meV due to an
acoustic phonon relaxation bottleneck. \cite{OrtnerPRB05} When
addressing arrays, this splitting can be resolved in the
inhomogeneously broadened spectra only for the 5 nm barrier
sample. For the samples with wider barriers it is too small to be
resolved. Also for these samples, no significant variations of the
exciton decay time are observed over the range of emission
energies. This demonstrates, that the lifetimes of the
electron-hole complexes are the same in the two states.

The decay is rather smooth for all sample, except for some wiggles
around 0.9 ns, where the sensitivity of the streak camera is
reduced. Some additional weak wiggles appear in the decay of the
molecules which might reflect variations of the exciton lifetime
at a given energy, in particular for the wide barrier samples.
They do not show any systematic dependence when varying the
detection energy (as can be seen from the contour plots in Fig.
1). When averaging over a finite energy range they are smeared
out, and will not be considered further in the following.

\begin{figure}
  \centering
  \centerline{\psfig{figure=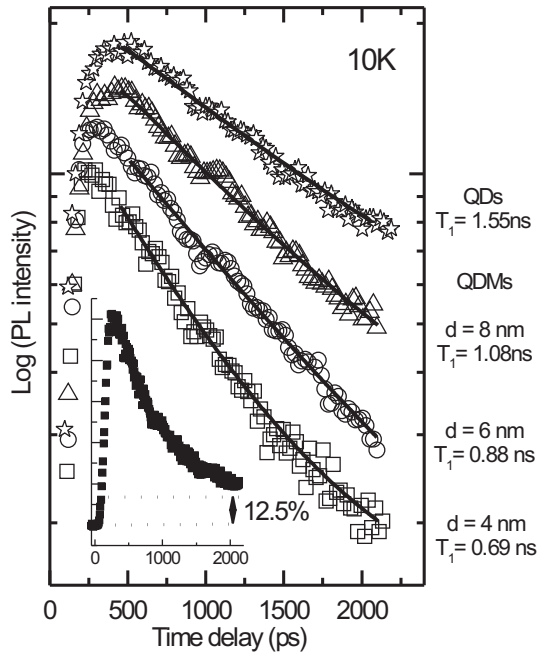,width=8truecm}}
\caption{\label{TRPLspectra}Photoluminescence decay traces for the
QDs reference and the 8, 6, and 4 nm barrier QDMs samples on a
logarithmic scale (vertically shifted for clarity), each taken at
energies in the center of the inhomogeneously broadened emission
peaks. Symbols give experimental data, lines are monoexponential
fits (see the text for details). The inset shows the $d$ = 4 nm
data on a linear scale to stress the existence of a constant
background of roughly constant intensity (the dotted line).}
\end{figure}

For better comparison, Fig. \ref{TRPLspectra} shows decay curves
(the symbols) taken at energies in the center of the emission
bands of QDMs with 4, 6 and 8 nm wide barriers and from the QDs
reference as a function of delay time ($T \sim$ 10 K). For better
visualization the traces have been shifted vertically. The
intensity has been normalized and is given on a logarithmic scale.
The samples differ slightly in the luminescence rise time (see
below) which is short as compared to the decay time, for which the
differences are much more pronounced. To determine the decay form,
the spectra have to be considered in more detail: after rise of
the signal all traces seem to follow an exponential dependence at
short delays. For longer delays deviations occur, as can be seen,
for example, from the trace for the $d$ = 4 nm sample, which is
shown in the inset of Fig. \ref{TRPLspectra} on a linear scale.
Aside from the fast decay there is a slow component, which appears
to be almost constant within the time range of interest.

As the optical excitation was done non-resonantly with a linearly
polarized laser pulse, we expect a quick depolarization of the
photoexcited carriers and in particular of the holes, since spin
relaxation times above barrier are rather short
\cite{MariePRB2000}. After relaxation spin-bright and spin-dark
excitons in the ground state therefore contribute to the decay
dynamics. \cite{background} The bright states will determine the
short decay behavior, while dark states can radiate only after a
spin-flip process. Generally a double exponential behavior is
therefore expected. The typical spin-flip time for holes is
considerably shorter than for electrons, but should still exceed
10 ns at low temperatures in quantum dots \cite{WoodsPRB2002}. The
dark excitons therefore will appear as a background that is almost
constant during the times considered here. From the data for the 4
nm barrier sample, we find that this constant background is 12.5
\% of the maximum signal strength, as obtained from an
extrapolation of the data toward long times. When subtracting this
background, the luminescence kinetics indeed follows to a good
approximation a single exponential decay.

To keep the number of fit parameters minimal, we also assumed the
same dark exciton background for the other samples, so that the
decay curves were analyzed by a fit function of form: $c_1
\exp(-t/T_1) + c_0$ with $c_0 = 0.125$. The solid lines in Fig.
\ref{TRPLspectra} give the resulting fits, from which reasonable
agreement with the experiment is seen. The decay times $T_1$
obtained by these fits are indicated at each trace. For the QDs
sample a relatively slow decay of 1.55 $\pm$ 0.1 ns is observed,
which can be understood from the small dot volume given by the
dimensions of 20 nm diameter and 2 nm height. Due to the
three-dimensional confinement the exciton wave function is
extended in momentum space with a width that can be estimated by 1
/ dot size. Only the components within the light cone can couple
to the light field and decay radiatively. The small dot size leads
to a considerable wave function spread beyond the light cone which
results in a long $T_1$ time. For the molecule samples $T_1$ is
strongly reduced as compared to this value. For the 8 nm barrier
QDM sample $T_1$ is 1.08~$\pm$~0.1 ns, and for the 4 nm barrier
sample $T_1$ is 0.69~$\pm$~0.1 ns, about a factor 2 smaller than
for the QDs. \cite{fittingdetail1}

\begin{figure}
  \centering
  \centerline{\psfig{figure=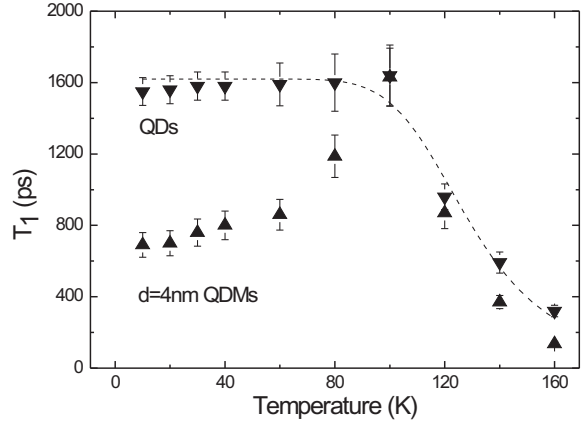,width=\columnwidth}} \caption{\label{Temperature}Temperature
dependence of the decay time $T_1$ from the QDMs sample with a 4
nm wide barrier (the up triangle symbols) and the QDs reference
sample (the down triangle symbols). The line is a guide for the
eye and follows equation (1) given in the text.}
\end{figure}

Before discussing the dependence of $T_1$ on the barrier thickness
in more detail we briefly report on its temperature behavior. Fig.
\ref{Temperature} shows the evolution of exciton lifetime with
increasing temperature from 10K to 160K for the QDs and the QDMs
with 4 nm barrier. The other QDMs samples have temperature
dependencies similar to the $d$ = 4 nm sample (not shown). In both
cases the decay time is about constant up to temperatures of 60 K.
This is known as a signature of the quasi-zero dimensional
confinement of excitons whose thermalization in k-space is
prohibited due to a delta like density of states \cite{YuAPL96},
preventing the charactristic linear (square-root) temperature
dependence of $T_1$ in quantum wells (quantum wires) to develop
\cite{FeldmannPRL87,AkiyamaPRL94}. For the QDs the lifetime does
not change up to even higher temperatures, while for the molecules
a considerable increase is observed, which is probably due to
thermal recycling of the carriers, that is, repeated excitation
into higher lying molecule states with subsequent relaxation into
the ground state. It is not fully clear why it appears for the
molecules and not for the dots. It could arise from reduced level
splitting, as the energy spacing between the tunnel split states
for all QDM barrier widths is smaller than the splitting between
the QD p- and s-shells.

We note that the integrated PL intensity does not change in the
temperature range up to 100 K, and the optical output is
comparable for the QD and the QDM samples. From this observation
we derive that non-radiative decay channels are of negligible
importance in these samples, so that the population lifetime can
be assumed to be equal to the radiative lifetime.

At higher temperatures above 100K a strong decrease of $T_1$
occurs which is accompanied by a quenching of photoluminescence.
We ascribe this regime to thermal emisssion of carriers out of the
dots. Taking into account a constant radiative recombination rate
1/$T_{rad}$ (equal to $1/T_1$ at low $T$) and a thermal emission
rate $\exp(-E_{A}/kT$)/$T_{esc}$ where $E_{A}$ and $T_{esc}$ are
the activation energy and the effective escape time, respectively,
one can fit the QD data according to the equation
\cite{thermalemission}
\begin{equation}
T_1 = \frac{T_{rad}}{1+ \frac{T_{rad}}{T_{esc}} \exp \left(-
\frac{E_{A}}{kT} \right)},
\end{equation}
A value $E_{A}$ = 95 $\pm$ 10 meV is obtained which is close to
half the barrier height between the molecule ground states and the
GaAs barrier. Such a result is expected for electron-hole pairs
from Boltzmann statistics assuming thermal equilibrium between
dots and surrounding. \cite{thermalemission}

\begin{figure}
  \centering
  \centerline{\psfig{figure=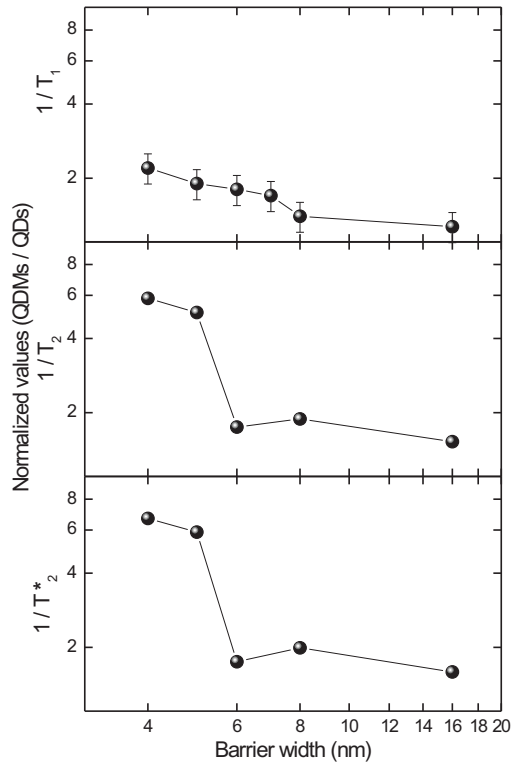,width=8truecm}}
\caption{Dependence of different decay rates on the QDM barrier
width normalized by the corresponding rate of the QDs reference
(plotted on a log-log scale): The upper panel gives the exciton
population decay rate 1/$T_1$ (error bars include the errors when
evaluating $T_1$ in QDMs and QDs), and the mid panel gives the
zero temperature exciton dephasing rate 1/$T_2$ for $T \rightarrow
0$ from ref. \cite{BorriPRL03}. From these quantities the pure
dephasing rate 1/$T_2^{\star}$ is calculated according to eq. (2)
and shown in the lower panel. For comparison all the rates are
shown on the same vertical  scale.} \label{TRPLrates}
\end{figure}

The upper panel in Fig. \ref{TRPLrates} shows the PL decay rates
at low $T$ as function of the molecule barrier width. The values
displayed are normalized by the corresponding rate of the QDs
reference. For all molecule samples, also for the wide barrier
cases, $T_1^{-1}$ is considerably larger than for the dot
structures. The enhancement is increasing with decreasing barrier
width. The radiative lifetime is inversely proportional to the
oscillator strength of an exciton, which is determined by the
overlap of the electron and hole wavefunctions and by its
coherence volume. In QDs the coherence volume is closely related
to the geometrical dot size \cite{AndreaniPRB1999}. Simply
speaking, the exciton collects oscillator strength from every
crystal unit cell into which its wave function spreads.

To understand the increase of the decay rate in the molecules, we
have to consider how quantum coherent coupling of two QDs affects
the oscillator strength. This problem was addressed theoretically
by Bryant \cite{BryantPRB93} for the case of laterally confined
excitons placed in symmetrically coupled quantum wells: The ground
state of an exciton in a QDM is a coherent superposition of
excitons in both QDs with twice the oscillator strength of one
exciton in a single QD. As a consequence we expect a reduction (an
increase) of the exciton lifetime (the decay rate) by a factor of
2.  The experimental value for the decay rate of 2.2$\pm$ 0.3
obtained for the QDMs with the narrowest barrier of 4 nm agrees
with this prediction, giving additional proof for an exciton wave
function that is extended over the molecule structure, for which
the two QDs must be coherently coupled. \cite{narrowbarrier}
Different penetration of electron and hole wave functions into the
barrier separating the dots might lead to a reduction of
oscillator strength, which is, however, expected to be
considerably smaller than the effect from doubling the coherence
volume. \cite{comment-ehoverlap}

Based on these arguments, however, we would expect a decay rate
increase independent of barrier width, provided coherent coupling
is established in a QDM. Instead a rather smooth increase with
decreasing barrier width is observed, starting from the 8 nm
sample with a ratio of about 1.4. Previous studies indicate that
coupling is established also for the wide barrier molecules.
\cite{OrtnerPRL03,OrtnerPRB05} Here two points have to be noted:
a) The measured decay times represent averaged values obtained
from ensemble studies. b) The doubling of oscillator strength is
expected for the case of symmetric molecules only.

The reduced decay rate value most probably arises from deviations
from ideal molecule symmetry resulting in variations of the
interdot coupling: Asymmetries arise from differences of the two
dots that form the molecule regarding geometry and composition or
from displacement of the dots relative to each other. Such
asymmetries were identified as origin of the complicated exciton
fine structure patterns reported in Ref. \cite{OrtnerPRL03} on
similar QDMs. From these studies we find a considerable variation
of the molecule coupling in the samples with wide barriers of 7
and 8 nm. Even though it is difficult to develop a criterion for a
statistical overview, the variety of fine structure magnetic field
dispersions obtained on many single QDMs suggests that about 40\%
of the molecules show asymmetries that are so weak that the fine
structure is identical to the one of an ideal molecule. For
another 40\% of the structures the asymmetries result in
anticrossings in the field dispersion, demonstrating quantum
mechanical coupling of reduced strength. For the final fraction of
molecules, the fine structure data do not permit us to make a
clear assignment as to whether the dots are coupled or not, but at
least the coupling is strongly reduced. When decreasing the
barrier width, the effects of asymmetries become less and less
important, as is expected from the increased tunnel splittings.
The fine structure data for the 4 and 5 nm barrier samples suggest
that all of the molecules are strongly coupled.

Let us discuss now the effect of these variations of molecule
coupling on the exciton oscillator strength.  The case of ideal
symmetry has been discussed already above. \cite{BryantPRB93} Any
symmetry reduction will reduce also the oscillator strength. To
give two examples: The tunnel coupling is basically suppressed,
when the dot asymmetry is bigger then the energy gain through
formation of a bonding orbital. In this case of two decoupled dots
one expects the same exciton decay rate as for the QDs reference.
If on the other hand the electron states are tunnel coupled but
the holes are localized because of smaller tunnel matrix elements
due to larger mass, the overlap of the carrier wavefunctions is
strongly reduced leading to exciton decay rates even smaller than
in the QD case.

Summarizing these discussions, the barrier width dependence of the
decay rate becomes understandable. For the strongly coupled
systems with 4 and 5 nm barrier, the effects of asymmetry are
negligible and a lifetime reduction by a factor of about 2 is
expected, in reasonable agreement with experiment. The decay rate
measured for the wide barrier samples represents an statistical
average to which 'ideal' molecule structures contribute with an
increase of $T_1^{-1}$ by a factor of 2 as well as structures with
reduced coupling with a $T_1^{-1}$ below 2, that could be even
smaller than the value for the QDs. Nevertheless, the fact that
the decay rate is roughly a factor 1.4 larger than for the QDs
even for the widest barrier sample, shows that most of the dot
pairs are molecules. The smooth increase of $T_1^{-1}$ reflects
the decreasing importance of asymmetries on the molecule coupling,
i.e. an increasing fraction of structures in the array resembles
'ideal' molecules, as expected from the increased coupling
strength.

The effect of asymmetries has to some extent been simulated in the
calculations of Bryant, \cite{BryantPRB93} who studied the exciton
coherence in coupled dots also under the influence of an electric
field along the molecule axis. Application of a bias leads to
carrier localization in one of the dots, and thus has similar
consequences as an geometry induced asymmetry of the molecules.
\cite{KrennerPRL05} The calculations show that the oscillator
strength resists weak bias changes due to the electron-hole
Coulomb correlations but for fields large enough to induce carrier
redistribution, there is a strong drop in oscillator strength, in
agreement with our qualitative arguments.

Now let us discuss, how the population decay rates $T_1^{-1}$
compare with the exciton dephasing rates $T_2^{-1}$ measured in
ref. \cite{BorriPRL03} at 10 K. These rates are connected through
the equation
\begin{eqnarray}
\frac{1}{T_2} = \frac{1}{2T_1} + \frac{1}{T_2^{\star}},
\end{eqnarray}
where $(T_2^{\star})^{-1}$ is the pure dephasing rate without
change of state population, for example due to relaxation between
fine structure states, virtual transitions involving phonons etc.
The mid panel of Fig. \ref{TRPLrates} gives the barrier width
dependence of $T_2^{-1}$ for the QDMs from ref. \cite{BorriPRL03}.
Given are the $T_2^{-1}$ that are obtained from extrapolation of
the four-wave-mixing data to zero temperature and are about
constant below 20 K. \cite{BorriPRL03} Again the data are
normalized by the corresponding rate of the QDs reference. As
mentioned, it has been shown that for QDs the dephasing is
ultimately limited by the radiative decay of the excitons.
\cite{LangbeinPRB04}

For the QDs reference we obtained a radiative decay time of $\sim$
1.6 ns, from which a dephasing time of more than 3 ns would be
expected in absence of pure dephasing. However, the experimentally
measured dephasing time is $\sim$ 600 ps only, showing that the
dephasing is not radiatively limited for the QDs under study but
pure dephasing plays an important role. The origin of this pure
dephasing cannot be clearly assessed from the present experiments.
Similar observations had already been made on other QD samples.
\cite{BorriPRL01} In general, the width of the zero-phonon line at
low temperatures which results from $T_2^{-1}$ is determined by
virtual phonon transitions (assuming negligible phonon
linewidths), \cite{MuljarovPRL04} whereas real phonon-assisted
transitions can be neglected due to a confined level splitting
that is large as compared to the thermal energy $k_B T$. These
virtual transitions might vary with the dots under study and might
provide an explanation for the difference between coherence and
population decay rates in some QD samples. Also a finite phonon
linewidth might become relevant for explaining the pure dephasing
(see below).

Similar to the population decay rate, also for the dephasing rate
an increase with decreasing barrier width has been observed in the
molecules. However, this increase is much stronger than for the
QDs. For the narrow barrier sample it reaches a factor of almost
6, as compared to the doubling of  $T_1^{-1}$. Hence the
difference between coherence and population decay rates becomes
even bigger for the molecules as compared to the dots. This can be
seen from the bottom panel of Fig. \ref{TRPLrates}, which gives
the pure dephasing rates normalized by the corresponding QDs
reference value. Vice versa, the exciton superradiance in the
molecules contributes only partly to the observed enhanced
dephasing. Mechanisms for pure dephasing also become more
important in the molecules. For example, the cross section for
virtual scattering processes might become larger. Further, one can
imagine that also real phonon assisted transitions play a role. In
the molecules the level splitting is reduced due to the tunnel
splitting of each state. This splitting is smaller than the QD
state splitting for all barrier widths. In particular for the hole
small splittings are expected, so that real transitions might
occur even at low $T$.

An alternate explanation might be provided by the recent
observation of a finite zero phonon linewidth that depends on the
dot surrounding. In single QD luminescence on laterally patterned
samples it has been observed that the enhancement of linewidth
with temperature is increased with decreasing size of the mesa
structures in which the dots were isolated for spectroscopy. By
detailed comparison with calculations, this dependence was traced
to exciton scattering with phonons, whose linewidth is determined
by the lateral surface roughness. \cite{OrtnerPRB04} This result
underlines the importance of interfaces in the surrounding of the
QDs on the exciton dephasing. As compared to dot structures, the
QDMs are subject to further interfaces due to the implemented
barrier. Certainly these interfaces will have roughness, due to
which the phonon linewidth might be increased as compared to the
QD case. \cite{dephasingannealing} With decreasing barrier width
this factor could become increasingly important due to penetration
of carriers into the barrier.

Indications that the tunnel splitting alters considerably the
carrier-phonon dynamics, are also found in the rise of the
photoluminescence signal. The rise time is $\approx$ 200 ps for
the QDs. A rise time of comparable length is observed for the wide
barrier molecules, while for the narrow barrier QDMs it is
significantly faster ($< 120$ ps). The rise is determined by the
relaxation of carriers from above the GaAs barrier into the ground
states of the confined geometries. For excitation into the wetting
layer, basically the same times were observed, so that the rise is
obviously dominated by the relaxation within the confined
geometries. In the wide barrier samples, the tunnel induced
splitting is quite small ($\approx$ 10 meV), so that the different
shells are still energetically well separated. Therefore no large
modification of relaxation is expected as compared to the dots.
For the narrow barrier samples, the tunnel induced splitting
becomes of the same order of magnitude as the shell splitting.
Thus the discrete QD spectrum is transformed by the molecule
formation into a spectrum where the states lie more densely on the
energy axis, potentially facilitating carrier relaxation.

In conclusion, we have demonstrated exciton superradiance induced
by coherent tunnel coupling in self-assembled InAs/GaAs QDMs and
have clarified its relation to exciton dephasing. To obtain more
detailed insight it would be interesting to study two-dot
molecules, in which the carrier distribution can be controlled on
a detailed level by application of an electric field along the
molecule axis. \cite{KrennerPRL05} It will be also interesting to
see whether more than two dot structures can be coherently coupled
to obtain a further enhancement of the exciton oscillator
strength. By placing such an arrangement into a high finesse
resonator, the regime of strong coupling of exciton and photon
with a sizeable Rabi-splitting among the hybridized states might
be entered. \cite{strongcoupling}

{\bf Acknowledgements} This work was supported by the Deutsche
Forschungsgemeinschaft (Forschergruppe 'Quantum Optics in
Semiconductor Nanostructures') and the Helmholtz-NRC research
foundation.


\begin{references}
\bibitem{nonanneal}
Details of the QDM fabrication can be found in Ref.
\cite{Fafard99}. For clarity we also note that the samples were
not thermally annealed.
\bibitem{singleQDM}
We note that we tried to perform also time resolved measurements
on single QDMs, to identify the effects of molecule geometry
variations. Unfortunately the sensitivity of our streak camera is
not sufficient to extract from the data reliable values for the
exciton decay times.
\bibitem{fittingdetail1}
Instead of considering a constant background we also tried a two
exponential fit but the time range of our spectra is too small to
get two distinct time scales.
\bibitem{narrowbarrier}
Due to intermixing the dots will contain also a considerable
fraction of Ga, while the barrier in-between will also contain In.
Therefore it i hard to identify sharp boundaries within the
molecule. The intermixing might facilitate a penetration of the
exciton into the barrier without reduction of the electron-hole
overlap, in particular for narrow barriers. \cite As the exciton
collects the oscillator strength from all crystal unit cells, into
which it spreads (see text), also the barrier might contribute to
the oscillator strength, so that it is enhanced by more than a
factor of 2 through the molecule formation.
\bibitem{comment-ehoverlap}
Studies of the Stark shift of excitons with a bias applied along
the growth axis show that in dome-shaped QDs the electron-hole
overlap is considerably reduced. For the In-flushed QDs with
heights of $\sim$ 2 nm we do not expect a strong displacement of
electron and hole. How this overlap is changed by the molecule
formation is not fully clear at the moment and requires further
studies. \cite{Fry}
\bibitem{dephasingannealing}
The roughness of the interfaces might be improved by thermal
annealing, which would reduce the importance of pure dephasing
$1/T_2^{\star}$. However, simultaneously the exciton coherence
volume will increase, leading to an increasing contribution of the
radiative decay rate $1/T_1$ to the dephasing rate $1/T_2$. Thus
we cannot expect a considerable prolongation of the QDM dephasing
times by annealing.
\end{references}
\end{document}